\documentclass[conference]{IEEEtran}
\usepackage{cite}
\usepackage{amsmath,amssymb,amsfonts}
\usepackage{algorithmic}
\usepackage{graphicx}
\usepackage{textcomp}
\usepackage{xcolor}
\def\BibTeX{{\rm B\kern-.05em{\sc i\kern-.025em b}\kern-.08em
    T\kern-.1667em\lower.7ex\hbox{E}\kern-.125emX}}
    
\usepackage{pgfplots,pgfplotstable}
\usetikzlibrary{patterns} 
\usepackage{graphicx}
\graphicspath{ {./figures/} }
\usepackage{adjustbox}
\usepackage{graphicx}
\usepackage{tabularx} 

\usepackage{booktabs,multirow,tabularx}

\begin{document}

\title{Memory-Latency-Accuracy Trade-offs for Continual Learning on a RISC-V Extreme-Edge Node}

\author{
  \IEEEauthorblockN{Leonardo Ravaglia\IEEEauthorrefmark{2}, Manuele Rusci\IEEEauthorrefmark{2}, Alessandro Capotondi\IEEEauthorrefmark{3}, Francesco Conti\IEEEauthorrefmark{2},\\ Lorenzo Pellegrini\IEEEauthorrefmark{1},
  Vincenzo Lomonaco\IEEEauthorrefmark{1}, Davide Maltoni\IEEEauthorrefmark{1}, Luca Benini\IEEEauthorrefmark{2}\IEEEauthorrefmark{4}}
  
  \IEEEauthorblockA{
    \IEEEauthorrefmark{1}DISI, University of Bologna, Italy\;\;
    \IEEEauthorrefmark{2}DEI, University of Bologna, Italy \\
    \IEEEauthorrefmark{3}FIM, University of Modena and Reggio Emilia, Italy\;\;
    \IEEEauthorrefmark{4}IIS, ETH Zurich, Switzerland
  }
  
}

\maketitle

\begin{abstract}
AI-powered edge devices currently lack the ability to adapt their embedded inference models to the ever-changing environment. 
To tackle this issue, Continual Learning (CL) strategies aim at incrementally improving the decision capabilities based on newly acquired data.
In this work, after quantifying memory and computational requirements of CL algorithms, we define a novel HW/SW \textit{extreme-edge} platform featuring a low power RISC-V octa-core cluster tailored for on-demand incremental learning over locally sensed data.
The presented multi-core HW/SW architecture achieves a peak performance of 2.21 and 1.70 MAC/cycle, respectively, when running forward and backward steps of the gradient descent.
We report the trade-off between memory footprint, latency, and accuracy for learning a new class with \textit{Latent Replay} CL when targeting an image classification task on the CORe50 dataset.
For a CL setting that retrains all the layers, taking 5h to learn a new class and achieving up to 77.3\% of precision, a more efficient solution retrains only part of the network, reaching an accuracy of 72.5\% with a memory requirement of 300~MB and a computation latency of 1.5 hours. On the other side, retraining only the last layer results in the fastest (867~ms) and less memory hungry (20~MB) solution but scoring 58\% on the CORe50 dataset. 
Thanks to the parallelism of the low-power cluster engine, our HW/SW platform results 25$\times$ faster than typical MCU device, on which CL is still impractical, and demonstrates an 11$\times$ gain in terms of energy consumption with respect to mobile-class solutions.\end{abstract}


\section{Introduction}
Novel sensors and smart devices are equipped with \textit{ultra-low-power} digital processing platforms that process raw data locally and extract high-level information by applying intelligent algorithms, such as Deep Neural Networks (DNNs).  
While the DNN inference capability has been already demonstrated on extreme-edge devices~\cite{zhou2019edge, chen2019deep, murshed1908machine}, the training of DNN models still relies on GPU-based machines. 
Once trained, the inference models are deployed on edge platforms tailored for prediction-only tasks. 
However, these inference models cannot adapt to the present environment, which may differ significantly from the training data statistics.

A way out of this rigid \textit{train-on-cloud -- deploy-at-edge} model has been proposed in the form of Continual Learning (CL) algorithms.
CL algorithms aim at adapting a network model to a new class of sensor stimuli. 
This process is extremely challenging because of the \textit{catastrophic forgetting}~\cite{serraovercoming}: learning new tasks or instances by fine-tuning only on new data leads to degraded recognition capabilities over initial data.
Safeguarding previous knowledge is, therefore, a significant concern. Regularization techniques and rehearsal-free methodologies, i.e., ones not reusing the initial training data, can tackle this problem; alternatively, rehearsal-based techniques safeguard previous knowledge with incremental training over a combination of previous and new data.
Among these latter techniques, Pellegrini et al.~\cite{pellegrini2019latent} presented a novel CL approach to effectively learn new object classes based on new samples and compressed old data, namely the Latent Replays (LRs). 

In this paper, we present a Hardware/Software platform design to run for the first time Continual Learning (CL) algorithms at the \textit{extreme-edge} on a MicroController Unit (MCU)-class architecture. 
In contrast with current \textit{ultra-low-power} MCUs that are mostly tailored for DNN inference, we propose a system architecture that can run incremental learning tasks on-demand by turning on a RISC-V-based 8-core cluster subsystem. 

The contributions of this paper are:
\begin{itemize}
    \item We evaluate the computational and memory requirements of a CL algorithm with latent-replays on the CORe50 NICv2-391 benchmark~\cite{lomonaco2019fine} with a MobileNetV1 model.
    \item We define a HW/SW architecture for CL at the \textit{extreme-edge} based on PULP~\cite{rossi2016pulp,pullini2018mr}, as well as a tensor tiling strategy necessary to fit within the limited memory.
    \item We benchmark forward and backward steps for CL on PULP, and evaluate their performance and efficiency  together with the energy-accuracy trade-offs of a MobileNetV1 network that learns over the CORe50 dataset.
\end{itemize}

The presented HW/SW design delivers an average performance of 1.84 MAC/cycle during the learning task, thanks to an almost ideal speed-up of 7.79$\times$ gained by the parallelization over 8-cores with respect to a single-core implementation.
Employing our platform, we demonstrate Continual Learning over the CORe50 dataset by coupling the processing engine with external memories for low-bandwidth operations. 
The learning of a new class can be achieved in 1.5~h by retraining only part of the network parameters with a memory need of 70~MB for RW operations and 200~MB to store old Latent Replay data permanently. This CL setting leads to an accuracy of 
72.5\%, which is only 5\% lower than retraining all the layers, but on it is 3.2$\times$ faster. 
Moreover, the proposed solution demonstrates to be 25$\times$ faster than typical low-power MCUs, and, given the estimated power cost of $\approx$70~mW, 11$\times$ more energy-efficient when compared to mobile-class solutions.
%

\section{Related Works}
\label{sec:related}

\subsection{Continual Learning}
The starting point of CL is \textit{Transfer Learning}~\cite{pan2009survey} that applies previously learned knowledge, i.e., the feature-extraction capability, on a different - but related - task. 
\textit{Incremental Learning}~\cite{losingaincremental} algorithms, instead, learn on new data to extend and enhance performances of a pre-trained model. 
On the other hand, incrementally training deep networks leads to catastrophic forgetting, in particular when learning over a stream of non-stationary data~\cite{french1999catastrophic,kemker2018measuring}. 
To tackle the forgetting problem \cite{serraovercoming}, a model can be fully retrained on a newly enriched dataset, including either old and new data. 
However, this latter approach results impractical on constrained embedded platforms, due to the severe limitations in terms of memory capability, that prevents the storage of a full dataset.

To face the forgetting challenge and keep the computational and memory load contained, novel CL algorithms learn over new data~\cite{kirkpatrick2017overcoming,lopez2017gradient, rebuffi2017icarl,lomonaco2019fine }. Among them, 
EWC~\cite{kirkpatrick2017overcoming} employs regularization terms inside the loss function to avoid catastrophic forgetting of previously learned samples. 
iCaRL~\cite{rebuffi2017icarl} is a class-incremental learner that makes use of a \textit{nearest-exemplar} algorithm for classification, so to exploit the similarity between classifiers to efficiently transfer previously learned knowledge. 
These algorithms differentiate on the update rules of the learning protocols and require relatively large batches for learning new classes but not yet solve the forgetting issues~\cite{parisi2019continual}.
The AR1 algorithm~\cite{lomonaco2019fine} partially solves the issues by retraining only a subset of the network layers based on the new data, to deal with memory and computational limitations of low-end devices, and making use of Batch Re-Normalization in place of Batch Normalization layers.   
Unfortunately, this approach may lead to a steep accuracy reduction (down to -20\% with respect to cumulative training on CORe50 NICv2–391~\cite{lomonaco2019fine}).
To deal with such degraded performance, Rehearsal-based techniques \cite{hayes2019memory, williams2019locally} keep in memory some representative patterns from past experiences.
During the incremental learning phase, the new frames are then interleaved with the past ones. 
In this context, a recent technique makes use of \textit{Latent Replays (LRs)}~\cite{pellegrini2019latent}, which are the activation feature vectors obtained by feeding the network with a subset of the training data. LRs are stored in memory, demanding a lower memory footprint w.r.t. RGB inputs.
This approach achieves state-of-the-art accuracy on the CORe50 NICv2-391 CL benchmark, recovering up to 15\% with respect to a Rehearsal-Free approach~\cite{pellegrini2019latent}. 
Due to the promising performance, we benchmark this latter CL algorithm  on the proposed HW/SW extreme-edge platform.

\subsection{Training at the Extreme Edge}
Edge and extreme-edge devices have constrained memory and computing resources that make the on-device training challenging.
Federated Learning addresses this limitation by distributing the training tasks on multiple devices ~\cite{mcmahan2017communication, konevcny2016federated}.
The work presented by Xu et al.~\cite{xu2019exploring} studied the trade-offs between computation latency for a training step and devices frequency
over six mobile-class processors while aggregating the results on the server side.
Some devices feature a higher frequency and power consumption with respect to others (up to 64\% higher) to balance the latency time among the different devices. 
Yet, the most efficient mobile-class platforms still present a power consumption in the range of few Watts. 
In contrast to them, we present an HW platform that allows training with a power budget below 100~mW.

From a platform-perspective, processing devices tailored for machine learning tasks feature a power consumption varying between a few mW up to hundreds of Watts~\cite{reuther2019survey}. 
We focus on flexible SW-programmable and low power devices distinguishing between 1-10~W (\textit{edge} device) or $<$1~W (\textit{extreme-edge} device) respectively.
Among the edge devices, we report the TPUEdge \cite{googleEdgeTPU}, which is developed mainly for embedded inference applications, 
and
the Qualcomm Snapdragon 845 that features a quad-core CPU, 4 Kryo 280 Gold coupled with an embedded GPU, and a dedicated DSP, namely the \textit{AI engine}. 
This latter device features a power consumption up to 4.5~W, which is above the requirement for an extreme-edge device that we target in this paper.

Moving to the \textit{ultra-low-power} spectrum, the \textit{extreme-edge} single-core MCUs, e.g., the STM32 MCU series~\cite{stm32h743ds}, feature low power consumption and compliant with the requirements of battery-powered sensors. 
However, this comes at the cost of the modest computational capacity, e.g. an STM32L4 device includes a single CPU running up to 48 MHz. 
To address this limitation, MrWolf \cite{pullini2018mr} features an MCU-based architecture accelerated by a multi-core cluster.
In particular, MrWolf has a power budget as little as 150~mW when running the 8-core cluster at 450~MHz and also supports floating-point (FP32) arithmetic. 
Relying on the design principle of this latter systems, we present a HW/SW design to enable CL at the \textit{extreme-edge}.

\section{CL with Latent Replays}
\label{sec:cl_model}

CL aims at fine-tuning deep networks based on new data samples, without retraining on the entire dataset.
Newly collected data used by CL can include new instances of previously seen classes, or instances of new classes, or both. 
This section sketchs \textit{i)} the computational model of the CL algorithm presented by Pellegrini et al.~\cite{pellegrini2019latent} when targeting an extreme-edge platform and \textit{ii)} compute its memory requirements.

\textbf{Computational Model}.
To bring CL with Latent Replay on a smart sensing platform, we model the incremental learning task to operate on a set of new data coming from a sensor, which is interfaced with an embedded digital processing engine. 
In the following, we focus on image classification as a representative problem for CL.

\begin{figure}[t]
    \centering
    \includegraphics[width=0.55\linewidth]{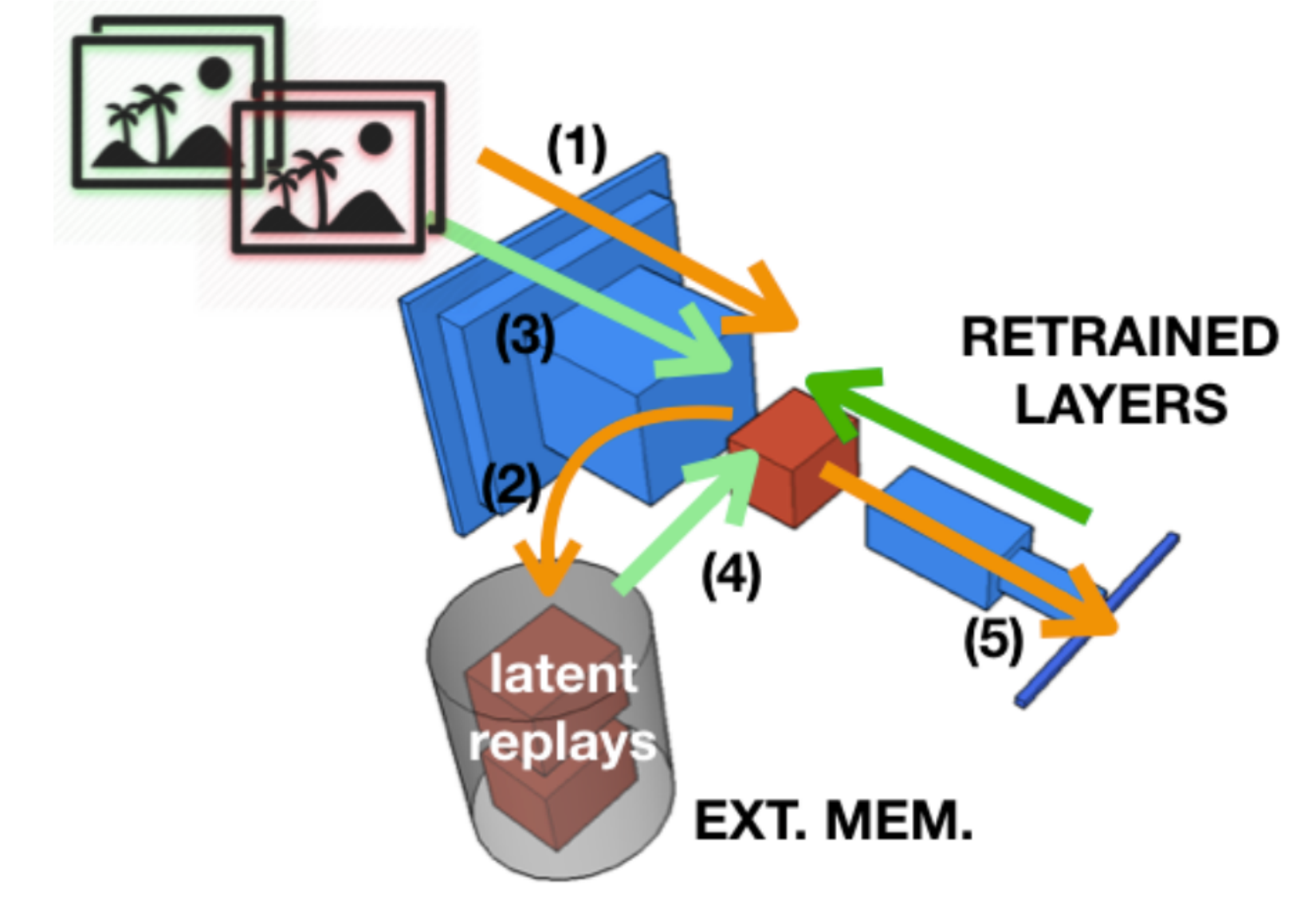}
    \caption{\textbf{Latent Replay computational model.}
   New images are elaborated up to the LR layer (1) and stored in the external memory (2). 
   Afterwards, a mix of new images (3) and LRs (4) are used to retrain the final layers (5).}
   \label{fig:latent-replay-figure}
   \vspace{-0.4cm}
\end{figure}

Fig.~\ref{fig:latent-replay-figure} illustrates the learning process that takes place over a set of data samples, including both $N_{I}$ new images and $N_{LR}$ old LR vectors, which are the feature maps of the Latent Replay ($LR$-th) layer, generated during previous trainings.

At runtime, a set of $N_{I}$ new class data are sampled and converted into LR vectors. 
For this purpose, the network is fed with the new data up to the Latent Replay layer. 
The new activations are then mixed with the old LR tensors, and afterwards, the learning algorithm updates the parameters of the remaining layers.
Hence, the dataset for a training step of CL  is composed of old and new LR vectors. 
Typically, $N_{I}$/$N_{LR} = 1/5$~\cite{pellegrini2019latent}. 
The batch gradient descent, which consists of forward and backward passes from the Latent Replay layer, is iterated for several epochs to converge to the best solution.
Furthermore, to face the catastrophic forgetting issue, we employ the AR1 training algorithm~\cite{lomonaco2019fine}.
Within the parameter update rule, AR1 applies a per-parameter scaling factor on the computed gradient, expressed by an approximation of the Fisher matrix. 
The Fisher matrix is defined as the integral of the loss function over the weights. 
The intuition behind this term is to keep the most meaningful parameters unchanged. 

\textbf{Memory Requirements}.
Consider a network model featuring $N$ stacked layers. 
Besides the data for the learning process, a total of $N_{w}=\sum_{i=0}^{N-1}~N_w(i)$ network parameters have to be stored in memory, where $N_w(i)$ represents the number of parameters at the $i$-th layer. 
Moreover, we account:
\begin{itemize}
    \item $N_{a}$ is the total amount of intermediate features computed during the forward pass, which have to be preserved in memory for gradient computation;
    \item $N_{g}$ is the number of gradient's components of the  network's parameters to be retrained.
    \item $N_{Fi}$ is the number of parameters in the Fisher matrix, which is equal to the number of parameters to update.
    \item $N_{fw}$ is the required memory footprint for temporarily storing activation feature maps during the forward pass. Its size is typically negligible with respect to other terms. 
\end{itemize}

\begin{figure}[t]
    \centering
    \includegraphics[width=0.8\linewidth]{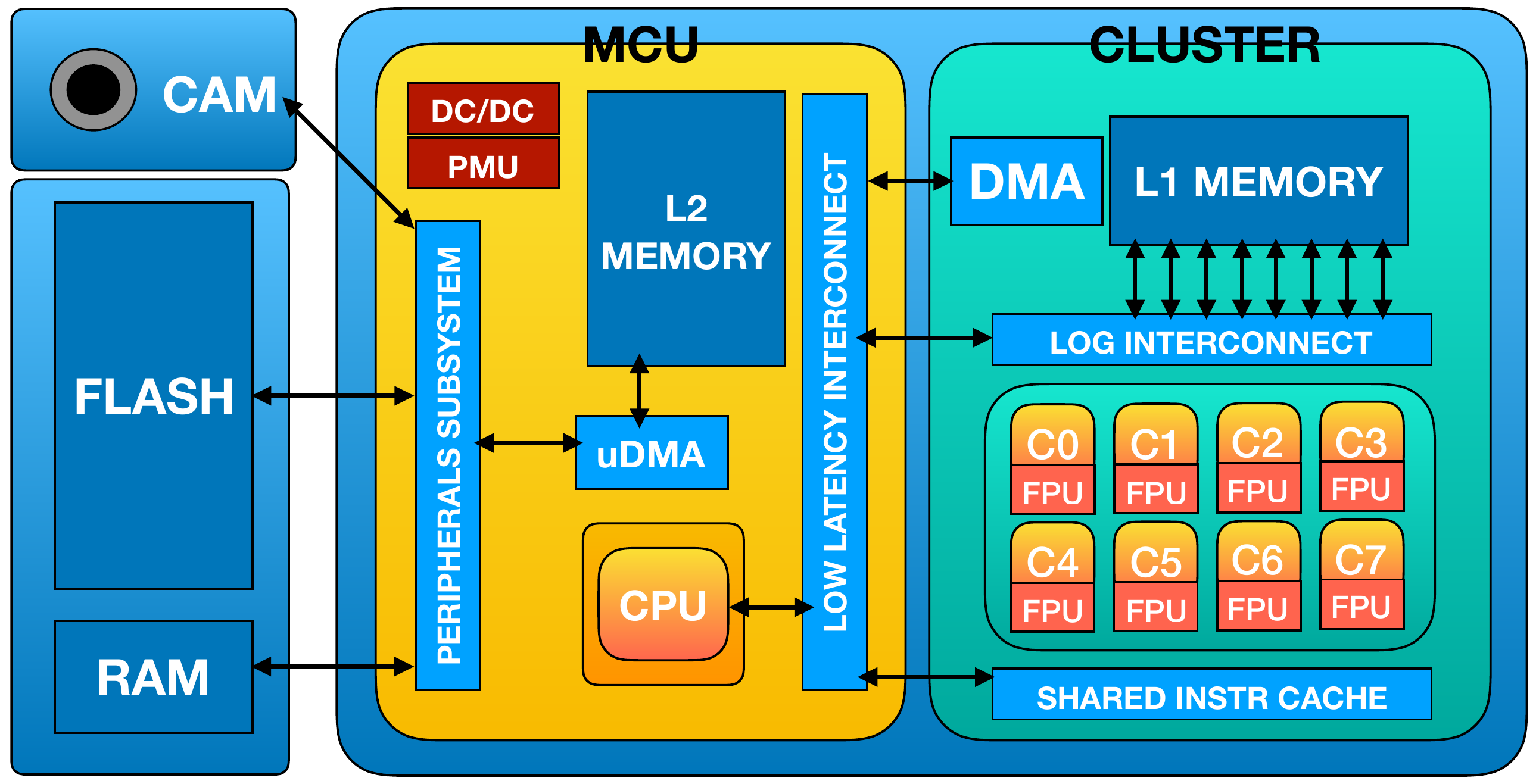}
    \caption{\textbf{The proposed platform architecture for \textit{extreme-edge} CL.} The architecture consists of a cluster of 8 RISC-V tightly coupled cores featuring private FPUs and two shared L1 and L2 scratchpad memories.}
    \label{fig:PULP-image}
    \vspace{-0.4cm}
\end{figure}

\section{HW/SW platform for CL}
In this section, we present our HW/SW platform design tailored for CL workloads. We evaluate the accuracy-latency-memory trade-offs based on the described system architecture.

\subsection{HW Platform for edge learning}

The proposed CL platform is based on the design principles of the PULP platform~\cite{gautschi2017near}, which combines parallel programming for high-performance and ultra-low-power features. 
The system architecture, which is depicted in Fig.~\ref{fig:PULP-image}, is based on an MCU platform accelerated by a multi-core cluster of RISC-V cores.
The MCU side features a single RISC-V core, namely the Fabric-Controller (FC), and a large set of peripherals. 
Besides the FC core, which is equipped with a private L1 data memory, the MCU-side of the platform includes a relatively large (256-1024~KB) on-chip L2 memory. 
The cluster features 8 processing elements (PE), each one including a RISC-V CPU equipped with a private Floating-Point Unit (FPU). 
To avoid memory coherency overheads and increasing area efficiency, the cores share a 64-256~KB L1 data scratchpad memory. 
Additionally, a multi-channel DMA engine handles data transfers between in-cluster and off-cluster memories. 

The L2 memory is used as a temporary buffer for the peripheral data: a $\mu$DMA unit autonomously handles the data transfers between the L2 memory and the external peripherals. 
Among them, the system can be interfaced with a large L3 off-chip memory, through a parallel interface, and an external FLASH for data storage through Quad-SPI, e.g., a SD FLASH memory. Both external memories feature low power consumption when in idle state. 

To efficiently implement power saving mechanisms, the FC core can switch-on the cluster on-demand at runtime, by controlling the internal cluster DC-DC regulator. 
Once powered-on, the FC core dispatches tasks on the 8-cores cluster, relying both on data- or task- parallelism for efficient and fast computation. 
On the other side, the system pays an energy overhead when activates the cluster computation on-demand.
Likewise, external memories can be power-gated if not running the learning tasks.

\subsection{SW stack for Continual Learning}
In this section, we detail the mapping of the CL data flows on the architecture defined in Fig.~\ref{fig:PULP-image}. 
Because of the high memory requirements and their constant nature, Latent Replays (LRs) are kept in external FLASH memory. 
During the learning phase, LRs are loaded into the L2 memory to feed the network model. 
On the other side, an external L3 DRAM memory is used to store network parameters, gradients, and intermediate activations values. 
Data transfers from L3 RAM to L2 occurs in the background of the computation thanks to the pipelined $\mu$DMA operation.

\begin{figure}[t]
    \centering
    \includegraphics[width=0.85\linewidth]{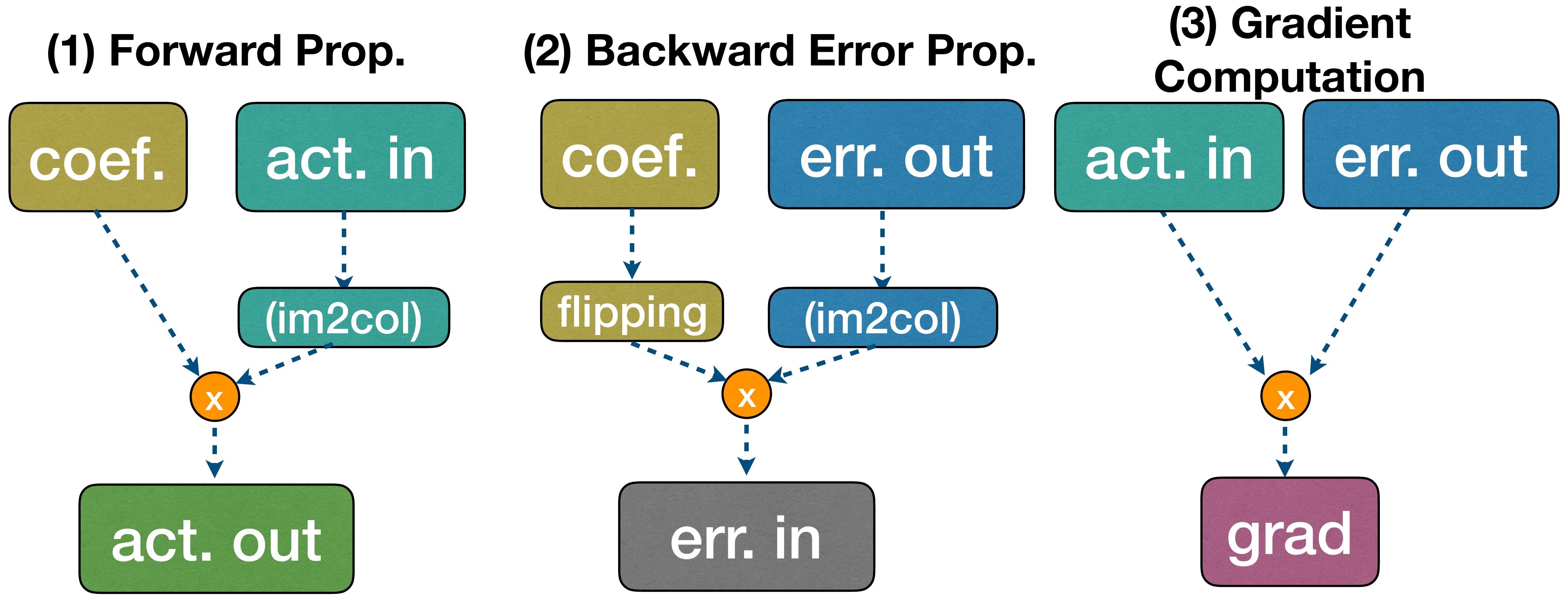}
    \caption{\textbf{CL Computational Phases Graphs.} (1) Forward step;  (2) Error back-propagation; (3) Gradient Descent computation.}
    \label{fig:graphs}
    \vspace{-0.3cm}
\end{figure}

\begin{figure}[t]
    \centering
    \includegraphics[width=0.83\linewidth]{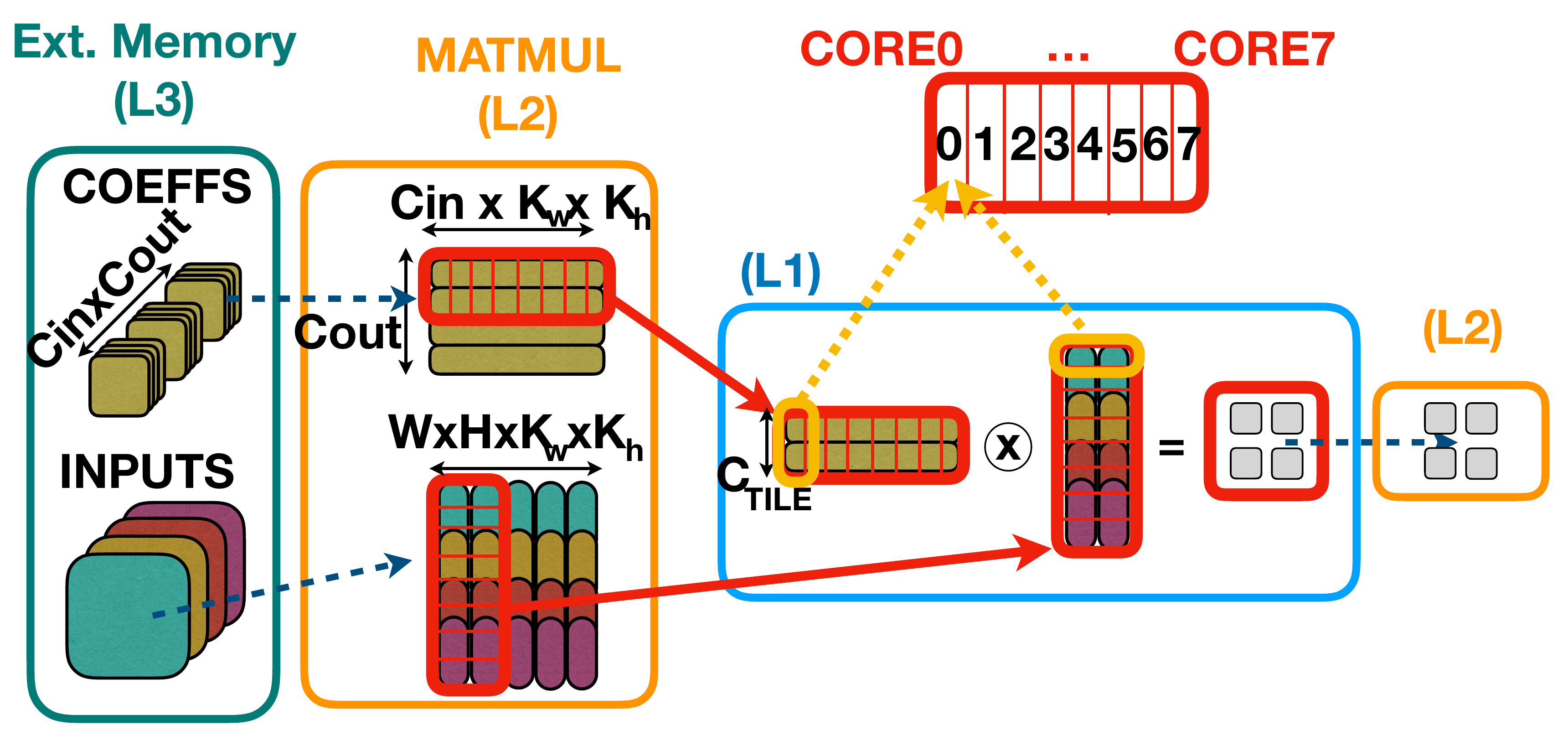}    
    \caption{\textbf{GEMM dataflow and parallelization scheme.} GEMMs use tiling to benefit from L2 and L1 data locality, and on GEMMs we use data-parallelism (SPMD) to distribute the work over the 8 cores of the cluster.}
    \label{fig:parallel-matmul}
    \vspace{-0.4cm}
\end{figure}

CL updates the model's parameters based on the back-propagation algorithm, which consists of the basic operations depicted in Fig.~\ref{fig:graphs}. 
During the \textit{Forward pass} of a network, the computational layers apply kernel convolutions over the activation values (\textit{act\_in}). 
Each convolution is reshaped as a  32-bit Floating-Point (FP32) generic matrix multiplication (GEMM), performed on the Cluster, by means of an \textit{im2col} transformation applied on the input activation tensor (\textit{act\_in}); the \textit{im2col} tensor is stored in a shared L2 buffer. 
The results of the forward kernels, i.e., the intermediate activation tensors, are then stored back in L3 memory for the update step. 
During the \textit{Backward Pass}, which runs up to the \textit{LR}-th layer, the activations gradient (\textit{err\_out}) is propagated to the next layer, after applying a convolution (GEMM) operation with the flipped (\textit{coeff}) vector (graph (2) in Fig. \ref{fig:graphs}). 
Likewise, the computation of the parameters gradient (\textit{grad}) is computed using a GEMM between \textit{act\_in}, computed during the forward pass, and the \textit{err\_out} tensors. 

Fig.~\ref{fig:parallel-matmul} represents the memory management and the parallelization scheme to implement the computational graph of Fig.~\ref{fig:graphs} in the target platform. 
Data (i.e., gradients, coefficients, activation values, or input data), which resides on the L3 memory, is loaded into the on-chip L2 and L1 memories for layer-wise processing.
Given the on-chip memory limitations, for some layers, tensor operands have to be sliced, and computation is performed on sub-tensors.
This tensor slicing is called \textit{tiling} and previous works already presented a lightweight strategy, hence implementing a software-cache mechanism for DNN deployment\cite{burrello2019work}.
In Fig. \ref{fig:parallel-matmul} we visualize the tiling of the coefficients along the output-feature dimension $C_{out}$: a subset of $C_{TILE}$ filters, hence a total of $C_{TILE} \times C_{in} \times K_{w} \times K_{h}$ parameters, is transferred to the cluster L1 memory for the computation.
Inside the cluster, the FP32 GEMM is \textit{parallelized} over 8 cores, both for forward and backward passes, according to a data-parallelism paradigm. 
Such scheme holds for the types of computations reported in Fig.\ref{fig:graphs}. The operand \textit{INPUTS} of Fig.~\ref{fig:parallel-matmul} refers either to the \textit{act\_in} or the \textit{err\_out} tensors. 
During forward propagation and gradient computation, the parallelization operates over the input-feature dimension $C_{in}$, whereas during backward error propagation the workload is parallelized over the $C_{out}$ dimension of the activations tensor. 

\section{Experimental Results}
In this section, we evaluate our HW/SW design over a \textit{Latent Replay} CL scenario, and we compare our solution against other proposed CL algorithm implementations. 

\subsection{Experimental Setup}

\textbf{CL Task and Dataset}.
To benchmark our HW/SW design, we consider a CL task consisting of learning a new object class from the CORe50 dataset~\cite{lomonaco2019fine}. 
This dataset includes 160k 128x128 training images of 50 objects.
To make the benchmark close to a real application, new objects are discovered sporadically over time. 
Thus, a three-way protocol was introduced (denoted as NICv2), where the first insertion of the samples of a new class is balanced over the training batches (see Fig.~3 in~ \cite{lomonaco2019fine}). 
In particular, in NICv2-391, each of the 390 incremental batches includes only one training session (300 images) of a single class.

We rely on the experimental settings of a previous study~\cite{pellegrini2019latent}, where the authors make use of a MobileNetV1 model with input resolution of 128x128 and width multiplier 1. In this section, we indicate the individual layers of the networks with the naming convention used by ~\cite{pellegrini2019latent} and also reported in Fig.~\ref{acc-mem-lat}.
The model is initially trained to distinguish only 10 of the 50 classes of the dataset. 
Each incremental learning step applies the gradient descent algorithm over a single batch composed of 1500 LRs vectors (30 LRs for every class) and 300 new images. 
This learning step iterates for 8 epochs. 

\textbf{Evaluation Metrics}.
The computation latency is measured through a cycle-accurate simulator of the proposed architecture. 
The base version of the simulator, which we
extended for this work, is validated against RTL for the open-source PULP platform and MrWolf, that is our reference SoC.
The selected running frequency is 150~MHz. 
The reported accuracy represents the overall precision reached by the network after complete learning on the CORe50 NICv2-391 benchmark. 
\subsection{Memory Evaluation}
\label{sec:memory}

\begin{figure}[t]
    \centering
    \includegraphics[width=0.83\linewidth]{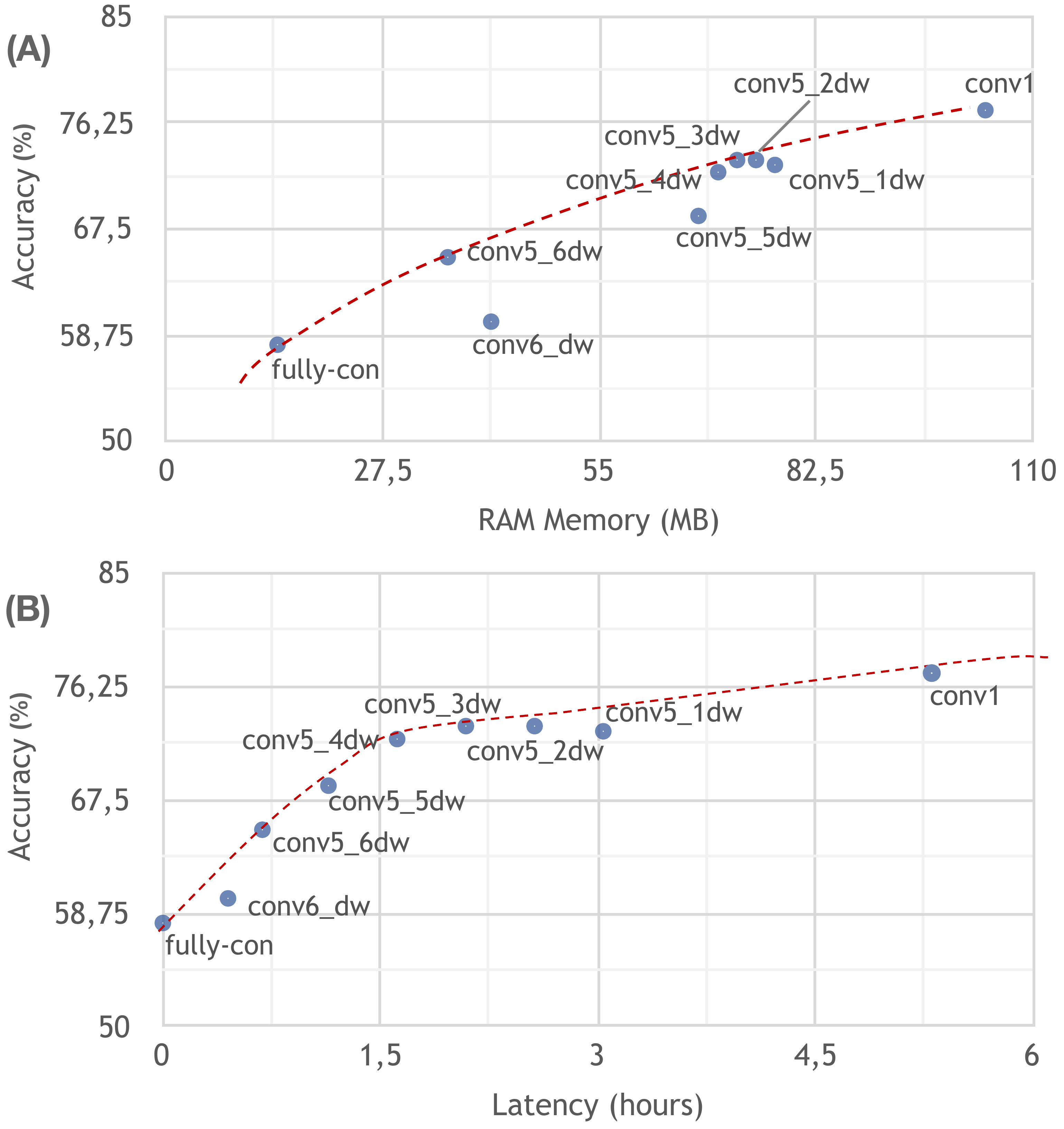}
    \caption{\textbf{Latency-Memory-Accuracy trade-off for different LR cuts.} The red dashed lines represent the two Pareto sets. The red dot (\textit{conv1}, the first layer) assumes that we retrain the whole network.
    }
    \label{acc-mem-lat}
    \vspace{-0.4cm}
\end{figure}

\begin{figure}[t]
    \centering
    \includegraphics[width=\linewidth]{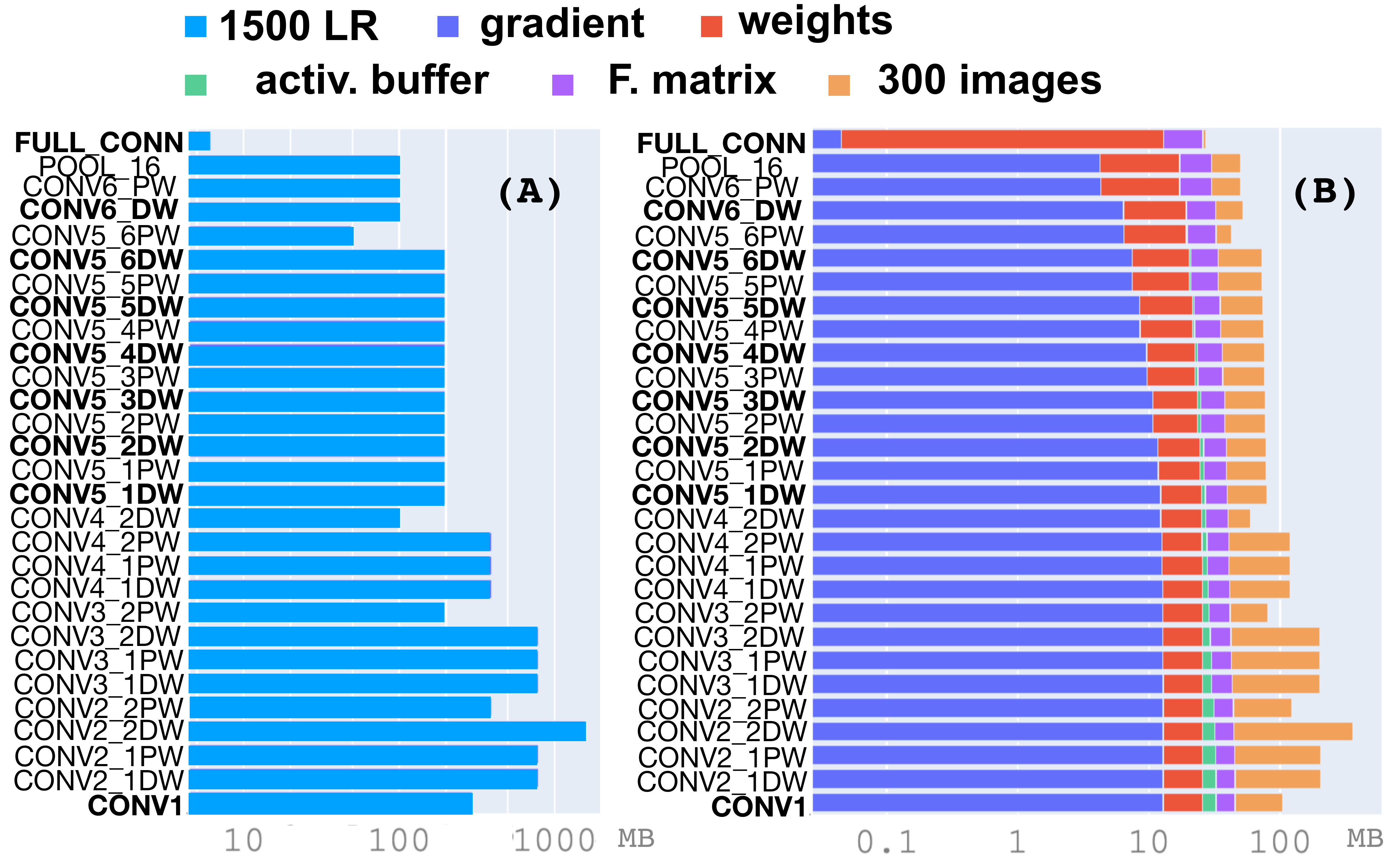}
    \caption{\textbf{FLASH and RAM Memory Footprint for different LR cuts.}}
    \label{flash}
    \vspace{-0.4cm}
\end{figure}

Fig.~\ref{flash} shows the memory footprint needed (A) to permanently store the 1500 LRs vectors, i.e. the ROM memory requirement, and (B) to store the new data samples, the network parameters and gradients, the intermediate features maps for the backward pass and the approximated Fisher matrix components. 
All these values result in a memory footprint of few MBs and change over time hence they are stored on the RAM.
Note that the memory requirements vary depending on on the chosen \textit{LR} layer: retraining only the last layer (indicated as $mid\_fc7$) requires the storage of smaller LR vectors and less gradients and activation features maps than selecting any other middle layer as the LR layer (e.g. $conv5\_4/dw$). 
This also explains the higher memory demands when choosing a LR layer closer to the first one.
Concerning the non-volatile FLASH memory (Fig.~\ref{flash}(A)), the requirement ranges from few MB (6~MB if the last layer is the LR layer) to up to 300~MB, when retraining the full network based on the image samples (the only CL setting not featuring LRs).  
A typical-size SPI FLASH memory is therefore sufficient in all cases.
Also note that LR arrays are accessed sporadically only to perform retraining.
Hence, the external FLASH memory can idle, or even be power-gated, for the rest of the time.

Fig.~\ref{flash}(B), instead, plots the RAM requirements, which also includes the 300 LRs related to the new images that require $>$60\% of the overall memory space.
The memory breakdown shows that individual memory terms vary depending on the LR layer selection. Only the size of network parameters is constant. 
Fig.~\ref{acc-mem-lat}(A) combines the RAM memory requirements with the network accuracy, 
which were demonstrated to be state-of-the-art on the CORe50 dataset~\cite{pellegrini2019latent}.
If setting the LR layer as one of the last layers,
the accuracy drop increases because of the higher number of frozen parameters.
If imposing a constraint of 32~MB for the external DRAM memory, the only match is the retraining of only the last layer (\textit{mid\_fc7}). 
But in this case, accuracy is limited to 58\%, which is nearly 20\% lower than the baseline (retraining all the layers). 
Conversely, to reach a higher precision of 72.2\% (\textit{conv5\_4/dw}), 70~MB of external RAM is needed, which is achieved using a multi-bank DRAM memory. 

\subsection{Latency-Accuracy Trade-off}

\begin{figure}[t]
    \centering
    \includegraphics[width=0.86\linewidth]{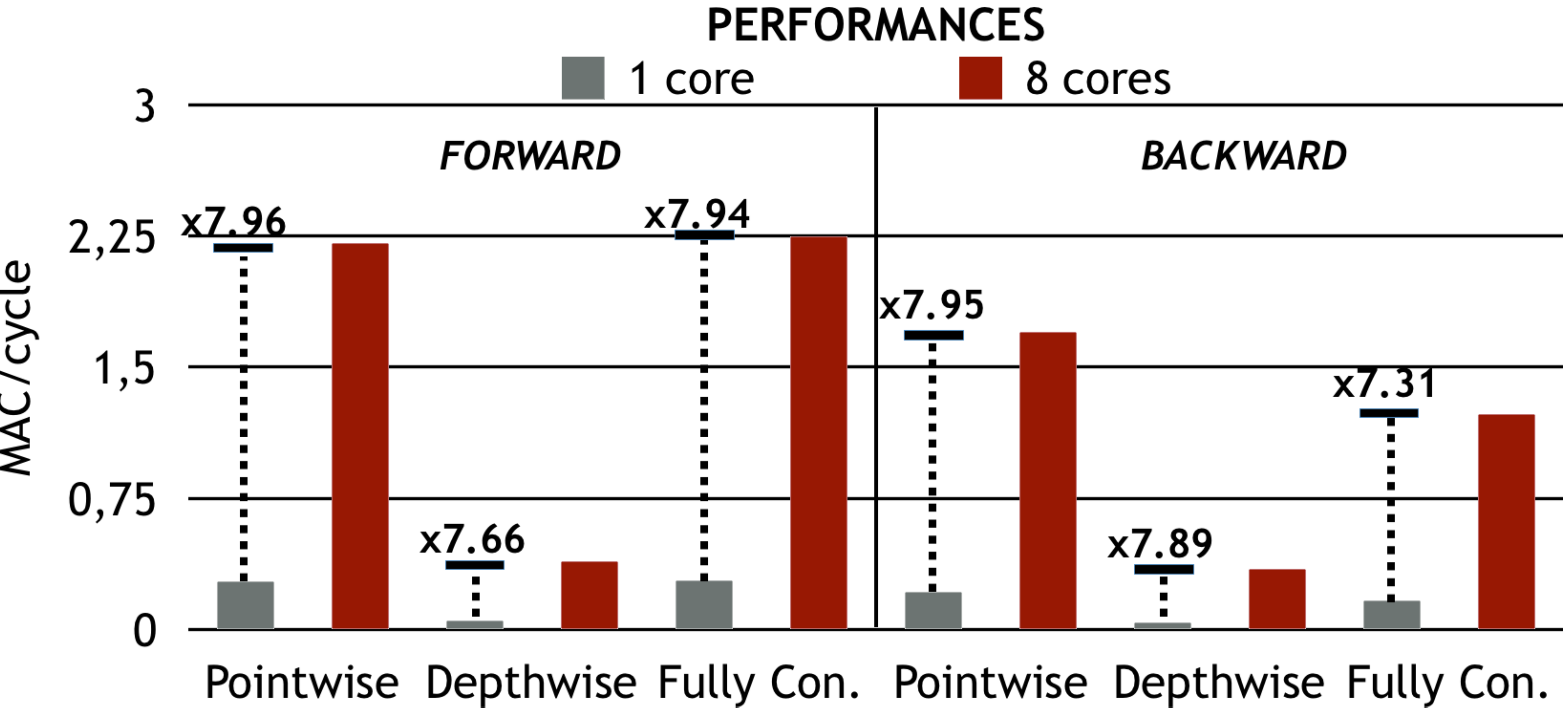}
    \caption{\textbf{Measured Forward and Backward Throughput and parallelization efficiency.}}
    \label{fig:performance}
    \vspace{-0.4cm}
\end{figure}

Fig.~\ref{fig:performance} reports the latency measured on our HW/SW platform over various computation kernels, either for single-core or 8-core execution.
The performance is expressed as MAC/cycle in case of forward and backward passes for three representative layers of our benchmark network, namely Pointwise, Depthwise, and Fully Connected layers. 
Peak values of performance are achieved in Pointwise layers, reaching 2.21 MAC/cycle in forward and 1.70 MAC/cycle in backward computations.
The performance measured for the forward pass results higher than backward because GEMM operates on larger vectors, hence the computation density is higher. 
The speed-up achieved by the 8-cores implementation against single-core reaches 7.79$\times$ on average, close to the theoretical limit of 8$\times$.

Fig.~\ref{acc-mem-lat}(B) reports the accuracy-latency trade-off on the learning task, with a cluster clock frequency of 150 MHz.
The latency refers to the total time to learn a new class, which depends on the chosen Latent Replay layer.
To reduce the memory access time overhead, we consider a tiling mechanism between L1 and L2 memories, realized using the DMA engine to minimize latency overhead (below 5\%~\cite{burrello2019work} with respect to the execution latency when on all data reside in L1). 
The L3 to L2 overhead instead is almost negligible because data transfers occur concurrently to the computation on large sub-tensors.
Retraining the whole network with 1500 replays results in an accuracy of $77.3\%$, and the overall latency for 1-class learning is of 318 min. 
A faster solution is obtained by choosing LR=\textit{conv5\_4}. In this setting, the learning latency for a new class is 98 minutes and the reached overall accuracy is 72.2\%. If we reduce the number of network's layers to retrain, the accuracy drops more substantially, as shown in Fig. \ref{acc-mem-lat}.
A \textit{ultra-fast} solution retrains only the Fully Connected layer, leading to a latency of 867~ms to learn a single class, but with a $58\%$ accuracy which sets a lower bound for the CL algorithm.
Our proposed platform opens up the possibility to perform CL on extreme-edge nodes, which is not practical on current-generation commercial  MCUs. For example, compared with a state-of-the-art low-power STM32L476 running at 48~MHz, our platform runs the learning task 25$\times$ faster.

\subsection{Energy Estimation}
We compare the energy consumption based on the CL mobile application presented in~\cite{pellegrini2019latent}, which operates over mini-batches of 500 replays and 100 new images, concluding the convenience and feasibility of CL at the extreme edge. 
The mobile-class device used for the reference design by Pellegrini et al. is a OnePlus6 featuring a Qualcomm Snapdragon 845 processor. 
To estimate the energy consumption of the proposed HW/SW platform, we refer to power measurements from the silicon prototype of MrWolf~\cite{pullini2018mr}, which operates at 9~MMAC/s/mW and features an average power of 70~mW when running at 150~MHz.

Given an application scenario requesting 1 inference/sec and 1 retraining step per hour, our platform features a total energy consumption of 34.2~J per hour in the fastest CL configuration setting (LR=\textit{mid\_fc7}). 
This implies a battery lifetime of about 710 hours (we consider a 3100 mAh battery), which is 11$\times$ higher than a Snapdragon845-based solution (assume $P_{idle}=0 W$).
To gain higher accuracy, hence retraining more layers, i.e., LR=\textit{conv5\_4} layer, the energy consumption increases to $1530$~J per hour, therefore leading to 15.8~h of battery life.

\section{Conclusions}

In this work, we presented a novel HW/SW architecture specifically tailored for the execution of CL algorithms. In particular we assess the memory and computational requirements of a Rehearsal-based CL algorithm with Latent Replay. 
Moreover, we show the accuracy-latency trade-off on the proposed \textit{extreme-edge} system, which results to be 11$\times$ more energy-efficient than previous mobile-class processors. 
The achieved results motivate to further explore Continual Learning on extreme-edge devices.

\bibliographystyle{IEEEtran}
\bibliography{sample-base} 

\end{document}